\documentclass[aps,amsmath,amssymb,twoside, showkeys,superscriptaddress, onecolumn,notitlepage, prd]{revtex4-2}
\pdfoutput=1
\usepackage[dvipsnames]{xcolor}
\usepackage{verbatim}
\usepackage{comment}
\usepackage{graphicx}
\usepackage[T1]{fontenc}
\usepackage{epsfig}
\usepackage{bm}
\usepackage{epstopdf}
\usepackage{amssymb}
\usepackage{float}
\usepackage{amsmath}
\usepackage{subfigure}
\usepackage{dcolumn}
\usepackage{hyperref}
\def\be{\begin{equation}}
\def\ee{\end{equation}}
\def\ba{\begin{eqnarray}}
\def\ea{\end{eqnarray}}

\def\doi{http://doi.org}

\newcommand{\HCd}{\mathcal{H}}

\def\HCdt0{\tilde{\HCd}_{0}}

\newcommand\mba{\mathbf{a}}
\newcommand\mbb{\mathbf{b}}
\newcommand\bee{\begin{equation*}}
\newcommand\eee{\end{equation*}}
\def\be{\begin{equation}}
\def\ee{\end{equation}}
\def\ba{\begin{eqnarray}}
\def\ea{\end{eqnarray}}

\newcommand\dphi{\dot{\phi}}
\newcommand\eb{b_0 e^{-\alpha\phi}}

\newcommand{\affbul}{Institute for Nuclear Research and Nuclear Energy, Bulgarian Academy of Sciences, Sofia, Bulgaria}

\begin{document}
\title{Special cases of the Multi-Measure Model -- understanding the prolonged inflation }
\author{Denitsa Staicova}
\email{dstaicova@inrne.bas.bg}
\affiliation{\affbul}

\begin{abstract}
The multi-measure model (MMM), in which one modifies the action to include both the Riemannian measure and a non-Riemannian one, has proven to be able to produce viable Universe evolution scenarios. 
In this article we consider two special cases of the multi-measure model, in which we first decouple the two kinetic terms in the Lagrangian entirely and later remove the dark charge of the model. We show numerically that those special cases still possess the needed evolutionary stages of the Universe and furthermore, for them one can obtain a sufficient number of e-folds of the early inflation. In the first case, the inflaton still moves backwards on the effective potential during inflation, while in the second, it does not, meaning that this behavior comes from the dark charge. We connect the model with hyperinflationary models and investigate how the different epochs are born from the interplay between the two scalar fields. We demonstrate that there is a dynamically induced slow-roll epoch, which is prolonged by the complicated movement of the two scalars in the field space. Finally, we show that while the adiabatic speed of sound can become imaginary, the phase speed of sound remains real.

\end{abstract}
\maketitle
\section{Introduction}
The main challenge in front of modern theoretical cosmology is not so much to construct an inflationary model as there is a long history of various such models (\cite{Guth:1980zm, Linde:1981mu, Linde:1984ir,  Liddle:1993fq, Damour:1997cb, 1402.0526, 1609.09781, Mukhanov:2005sc}). The challenge is to produce a model which satisfies all the observational requirements, while attempting to solve the known problems. In order for a theory to be considered viable, it needs to reproduce the main cosmological periods -- early inflation, matter domination and late-time exponential expansion and the graceful transition between them. It also should to be able to produce a powerful enough inflationary stage and not to contradict the predictions of the observational data \cite{Riess_1998, Riess_2019, Aghanim:2018eyx, Peiris:2003ff}. The latter means, it needs to reduce to the extensively tested $\Lambda-CDM$ model at post-inflationary times and eventually to adress some of the current tensions, like the tension in the measurements of the present day Hubble constant \cite{Escamilla_Rivera_2020,divalentino2020cosmology, benisty2020testing,Hryczuk_2020,Ivanov_2020, Alestas_2020}. 

The multi-measure model (MMM) has been introduced in series of works by Guendelman, Nissimov, and Pacheva \cite{hep-th/0106084,gr-qc/9905029, gr-qc/0211095, 1303.7267, 1205.1056, 1403.4199, 1407.6281, 1408.5344, 1609.06915, 1507.08878, 1603.06231}. Its main advantage is that in the case of one scalar field (the so called darkon), it possesses a dynamically generated cosmological constant. This means that it allows for spontaneous symmetry breaking, starting from Weyl (local conformal) invariant theory.  
This is important, because the Planck results predict almost scale invariant spectrum of primordial fluctuations in the CMB \cite{1807.06211}. It does not change the speed of light and thus it remains within the constraints set by LIGO \cite{Abbott_2017, Cornish_2017}. Finally, the models based on the non-Riemannian measure have been studied extensively in various physical situations and have shown promising results \cite{1905.09933, 1906.06691, 1907.07625, 1912.10412, 2002.04110, 2003.04723, 2003.13146,Guendelman:2022cop}.

The multi-measure model employs a number of scalar fields coupled to more than one independent volume form. In all the models, one of the volume forms is the standard Riemannian volume form, proportional to the square root of the metric determinant, and the other, non-Riemannian volume forms, can be defined trough the derivatives of auxiliary third rank anti-symmetric gauge field(s) (exact four-form). These new, auxiliary fields add only gauge degrees of freedom, but they dynamically generate a cosmological constant as a consequence of the equations of motion. Furthermore, they lead to a perfect fluid energy-momentum tensor, describing the dark energy and the dark matter sector simultaneously. 

In our previous work \cite{1610.08368, 1801.07133, 1906.08516}, we studied the cosmological aspect of MMM with two scalar fields -- the darkon and the inflaton. We demonstrated that it can describe well phenomenologically the evolution of the Universe, but we also noted some weakness. Notably, the model could not produce the needed number of e-folds. This seemed like a numerical problem, but due to the large number of parameters, we were not able to prove it. In this article, we continue our work on the model by considering two special cases -- first we remove the coupling between the Lagrangians of the two scalar fields and second, we remove the dark charge. This simplifies the problem and allows us to better study its parameter-space. We see that we are able to obtain the needed number of e-folds ($>60$) and that they depend strongly on the initial size of the universe. Also, we show numerically that there is a dynamically induced slow-roll period produced by the complicated movement of the two scalar fields which leads to the early inflation. Similar prolongation of the inflationary period has been observed in the so called hyperinflation models \cite{Pashitskii:2015ioa, Brown:2017osf, Mizuno:2017idt, Bjorkmo:2019aev,Bjorkmo:2019fls, Christodoulidis:2019mkj, Romano:2020oov, Ferreira:2020qkf}. Finally we investigate the speed of sound in the two cases and we show that as expected, there is a difference between the adiabatic and the effective speed of sound. 

\section{The Multi-Measures Model}
{The multi-measure model has been described in \cite{1609.06915, 1408.5344,1610.08368, 1801.07133, 1906.08516}. It  features  two   scalar fields --- an inflaton $\phi$ and darkon $u$ and its action is $S= S_{darkon}+S_{inflaton}$, with:}
\begin{align}
&S_{darkon}=\int{d^4x(\sqrt{-g}+\Phi(C))L},\notag\\
& S_{inflaton}=\int d^4x \Phi(A)(R+L^{(1)})+\int d^4x\Phi(B)\left(L^{(2)}+\frac{\Phi(H)}{\sqrt{-g}}\right).
\label{inflA}
\end{align}

{In this action, we have the Riemannian measure $\sqrt{-g}$, along with the non-Riemannian measures $\Phi(X)$ (where $X=A,B,C,H$), defined as generally covariant integration measure density dual to the field-strengths of an auxiliary 3-index antisymmetric tensor gauge field $X_{\nu\kappa\lambda}$ (i.e. $\Phi(X)=\frac{1}{3\!}\epsilon^{\mu\nu\kappa\lambda}\partial_\mu X_{\nu\kappa\lambda}$).}

{The Lagrangians are defined as:}
\begin{align}
& L=-\frac{1}{2}g^{\mu\nu}\partial_\mu u\partial_\nu u -W(u),\\
& L^{(1)}=-\frac{1}{2}g^{\mu\nu}\partial_\mu\phi\partial_\nu \phi- V(\phi),\; V(\phi)=f_1 e^{-\alpha \phi} \\
&L^{(2)}=-\frac{b_0}{2}e^{-\alpha\phi} g^{\mu\nu}\partial_\mu\phi\partial_\nu \phi + U(\phi),\; U(\phi)=f_2 e^{-2\alpha \phi}
\label{lagr}
\end{align}
\normalsize

{The form of the Lagrangians is chosen in such a way, so that the final total action $S$ is invariant under the following global Weyl scale transformation:}
\ba
&& g_{\mu\nu}\rightarrow \lambda g_{\mu\nu},\;\;
\Gamma^\rho_{\mu\nu}\rightarrow\Gamma^\rho_{\mu\nu},\;\;
\phi\rightarrow\phi+\frac{1}{\alpha}\ln\lambda,\;\;
u\rightarrow \frac{1}{\sqrt{\lambda}} u,
\notag\\
&&A_{\mu\nu\rho}\rightarrow \lambda A_{\mu\nu\rho},\;\;
B_{\mu\nu\rho}\rightarrow \lambda^2 B_{\mu\nu\rho},\;\;
C_{\mu\nu\rho}\rightarrow \lambda^2 C_{\mu\nu\rho},\;\;
H_{\mu\nu\rho}\rightarrow H_{\mu\nu\rho}
\label{winv}
\ea

{An interesting property of the so defined action is that the variation of $S$ with respect to the auxiliary fields $A, B, C$ and $H$ leads to four
 dynamically generated integration constants $M_0,\;M_1,\;M_2$ and $\chi_2$:}
 \ba
 L&=&-2M_0,\qquad R+L^{(1)}=-M_1, \nonumber\\
 \frac{\Phi(B)}{\sqrt{-g}}&=&\chi_2,\qquad L^{(2)}+\frac{\Phi(H)}{\sqrt{-g}}=-M_2.
 \label{const}
\ea

{In these equations, $M_1$ and $M_2$ are dimensionful and $\chi_2$ dimensionless constants, such that $\chi_2$ preserves global Weyl-scale invariance, while $M_1$ and $M_2$ leads to dynamical spontaneous breakdown of global Weyl-scale invariance under \ref{winv} due to the scale non-invariant solutions of Eq. \ref{const}.}

{After performing the variations with respect $\Gamma$, one can eliminate the auxiliary fields in the model leaving an effective Lagrangian depending only on $g_{\mu\nu}$ and the inflaton scalar field. This means that in Einstein frame, we have the standard general relativity action satisfying a perfect fluid energy-momentum tensor $S^{(eff)}=\int{d^4x \;\sqrt{-\tilde{g}}(\tilde{R}+L^{(eff)})}\label{sef}$  (for more details on the derivation see \cite{1906.08516}).}

{The effective Lagrangian in Einstein frame has the following form:}
\be
L^{(eff)}=\tilde{X}-\tilde{Y}(V(\phi)+M_1-\chi_2\eb\tilde{X})+\tilde{Y}^2(\chi_2(U(\phi)+M_2)-2M_0).
\label{l_b0}
\ee

Here, $\tilde{X}=-\frac{1}{2}\tilde{g}^{\mu\nu}\partial_\mu\phi \partial_\nu\phi 
$ and $\tilde{Y}=-\frac{1}{2}\tilde{g}^{\mu\nu}\partial_\mu\tilde{u} \partial_\nu\tilde{u}$,
are the respective kinetic terms for the two scalar fields in the Weyl-rescaled metric $\tilde{g}^{\mu\nu}$. The potential terms are $V(\phi)=f_1 e^{-\alpha \phi},\,  U(\phi)=f_2 e^{-2\alpha \phi}$ from eqs. \ref{lagr} and $M_0, M_1, M_2$ are the integration constants from eq. \ref{const}. The ratio $\chi_2=\Phi(B)/\sqrt{-g}$ is the only left-over from the non-Riemannian measures in the Lagrangian in Einstein frame.

This effective Lagrangian is non-linear, it has non-canonical  kinetic terms of both scalar fields and thus can be classified as a generalized k-essence type. We also have a coupling parameter $b_0$ between the two kinetic terms $\tilde{X}$ and $\tilde{Y}$. 

In the Friedman--Lemaitre--Robertson--Walker space-time metric, the effective equations of motion are: 
\ba
v^3+3\mba v+2\mbb&=&0 \label{sys1}\\
\dot{a}(t)-\sqrt{\frac{\rho}{6}}a(t)&=&0 \label{sys2}\\
\frac{d}{dt}\left( a(t)^3\dphi(1+\frac{\chi_2}{2}\eb v^2) \right)+
a(t)^3 (\alpha\frac{\dphi^2}{4}\chi_2\eb v^2+\frac{1}{2}V_\phi v^2-\chi_2 U_\phi\frac{v^4}{4})&=&0\label{sys3}
\ea

The dot over the fields indicates the time derivative and the subscript $\phi$ -- the derivative with respect to the field $\phi$. From here on we will omit writing the explicit dependence on $\phi$ of $V(\phi)$ and $U(\phi)$ and where it makes sense, the time-dependence of $v(t)$.

The algebraic Equation (\ref{sys1}) comes from the conservation of the dark charge where $v=\dot{u}$ and
the parameters are:
 $$\mba_{}=-\frac{1}{3}\frac{V+M_1-\frac{1}{2}\chi_2 b_0 e^{-\alpha\phi}\dot{\phi}^2}{\chi_2(U+M_2)-2M_0}, \mbb_{}=-\frac{p_u}{2a(t)^3(\chi_2(U+M_2)-2M_0)}$$ 
with $p_u$ --  an integration constant corresponding to the so called ``dark charge''.

Equation (\ref{sys2}) is the first Friedman equation where 
$a(t)$ is the metric scaling function, and the energy density is:
 $$\rho=\frac{1}{2}\dphi^2 (1+\frac{3}{4}\chi_2 b_0 e^{-\alpha\phi} v^2)+\frac{v^2}{4} (V+M_1)+
 \frac{3 p_u v}{4a(t)^3}. $$
 
 The second Friedman equation is: 
\begin{equation}
\ddot{a}(t)=-\frac{1}{12}(\rho+3p)a(t),\label{dda} 
\end{equation}
where the pressure of the perfect fluid is: $p=\frac{1}{2}\dphi^2 (1+\frac{1}{4}\chi_2\eb v^2)-\frac{1}{4}v^2(V+M_1)+p_u v/(4 a(t)^3)$.
 
\section{Special cases}
First we will consider the case $b_0=0$. This means that we are removing the coupling between the two kinetic terms in the effective Lagrangian so that it becomes: 
 \be
L^{(eff)}=\tilde{X}-\tilde{Y}(V+M_1)+\tilde{Y}^2(\chi_2(U+M_2)-2M_0).
\label{l_c0}
\ee

This is still a non-linear Lagrangian of the k-essence type with inflaton equation (Eq. \ref{sys3}) as follows: 

\ba
\ddot{\phi} + 3\dot{\phi}\frac{\dot{a}(t)}{ a(t)} -f_1\alpha e^{-\alpha\phi(t)}v(t)^2/2 + \chi_2 f_2 \alpha e^{-2\alpha\phi(t)} v(t)^4/4 = 0.
\ea
 
The velocity of the darkon scalar field $u$ becomes:

 \ba 
 v(t)=&\left(\frac{2U_{eff}}{V_M^2}\left (p_u/(a(t)^3 + \sqrt{-\frac{16}{81} U_{eff} V_M + p_u^2/a(t)^6}\right)\right)^\frac{1}{3}+\nonumber \\
 &\quad \quad \quad \frac{4U_{eff}}{3V_M}\left(\frac{2U_{eff}}{V_M^2}\left (p_u/(a(t)^3 + \sqrt{-\frac{16}{81} U_{eff} V_M + p_u^2/a(t)^6}\right)\right)^{-\frac{1}{3}}.
 \label{v_b0}
 \ea

 We recall that the effective potential of the theory is defined as:
 \begin{equation}
U_{eff}(\phi)=\frac{(f_1 e^{-\alpha\phi}+M_1)^2}{4\chi_2 (f_2 e^{-2\alpha\phi}+M_2)-8 M_0}.
\end{equation}
 
 and $V_M=f_1e^{-\alpha\phi(t)} + M_1$. 

From Eq. \ref{v_b0} we can easily see that there is an initial singularity in our equations connected with the term $1/a(t)^3$. Also, $v(t)$ may become complex, so it is important to work with such parameters for which it remains on the real plane. 

For $b_0=0$, the energy density becomes:  
 $$\rho=\frac{1}{2}\dphi^2 +\frac{v^2}{4} V_M+3 v \frac{p_u}{4a(t)^3}. $$
 
In view of the possible initial singularity for $a(t)=0$, it is useful to consider the asymptotics for $v(t)$. For $t\to0, a(t)\to0$, 
\ba 
 v(t)= \left(  \frac{2U_{eff}}{V_M^2} \frac{2 p_u}{a(t)^3} \right)^\frac{1}{3}+
 \frac{4U_{eff}}{3V_M}\left(\frac{2U_{eff}}{V_M^2}\frac{2p_u}{a(t)^3}  \right)^{-\frac{1}{3}}.
 \ea
 
Here the two terms have equal real parts but opposite imaginary parts so that  $v(t)$ remains real. Thus, one can assume that in this limit,  $v(t)\approx 2\Re\left (\left(  \frac{2U_{eff}}{V_M^2} \frac{2 p_u}{a(t)^3} \right)^\frac{1}{3}\right).$ 

If we use that value to find an approximation for the inflaton equation around the singularity at $t=0$, we find: 

\ba
\ddot{\phi} + 3\dot{\phi}H + W=0
\label{infl_sing}
\ea

\noindent where $W=2\chi_2 f_2 \alpha e^{-2\alpha \phi(t)} \left( \sqrt{2}  \frac{U_{eff}}{V_M^2} \frac{p_u}{a(t)^3} \right) ^{4/3}  - f_1\alpha e^{\alpha\phi(t)} \left( \sqrt{2} \frac{p_u}{a(t)^3} \frac{U_{eff}}{V_M^2}\right)^{2/3}$ and  $H=\frac{\dot{a}(t)}{ a(t)}$ is the Hubble constant and the prime denotes derivative with respect to $\phi$. This term is qualitatively different from $U_{eff}'$ due to the critical dependence on $a(t)$.  The density $\rho$ and the pressure $p$ also depend strongly on $a(t)$. This approximation is applicable only very close to $t=0$, under our numerical setup, until about $t\sim 10^{-3}$. 

Second, we study the case $b_0=0, p_u=0$. Setting $p_u=0$ corresponds to the asymptotic for $v(t)$ in which $a(t)$ is far away from the initial singularity: 

\ba
v(t)= 2\sqrt{\frac{U_{eff}}{V_M}} = \sqrt{\frac{(f_1 e^{-\alpha\phi(t)} + M1)}{\chi_2 (f_2e^{-2\alpha\phi(t)} + M_2) - 2 M_0)}}
\label{v_c0}
\ea

When $p_u\neq0$, i.e. in the first case, this approximation of the darkon velocity is excellent fit for the actual velocity for $t>10^{-3}$. Accounting for the much simpler form of $v(t)$, the inflaton equation becomes:
\ba
\ddot{\phi} + 3\dot{\phi}H + U_{eff}'=0.
\label{infl_eq}
\ea

This is the standard inflaton equation of a single scalar field rolling down a potential. In this case the density and the pressure become: $\rho=\dot{\phi}^2/2+U_{eff}$ and $p=\dot{\phi}^2/2-U_{eff}$, thus simplifying dramatically the Friedman equation. Therefore, the second case corresponds to a single scalar field case moving in a non-trivial effective potential, with an additional equation of state for the scalar field coming from the algebraic Eq. \ref{v_b0}. For these much simpler EOM, one can see that the equation of state of the universe (EOS) $w=p / \rho$ still satisfies the observational requirements ($ w^{a(t)\to0}\xrightarrow{} 1/3, w^{a(t)\to\infty} \xrightarrow{} -1$) analytically. 

{\bf Numerical methods}

To integrate numerically the system, we use the Fehlberg fourth - 
fifth order Runge--Kutta method with degree four interpolation implemented in Maple.  We perform our calculations in units in which $c=1$, $G=1/16\pi$, and $t_u=1$, where $c$ is the speed of light, $G$ is Newton's constant, and $t_u$ is the present day age of the Universe (i.e. we normalize all our solutions so that $a(1)=1$). We also normalize the matter-domination epoch to start at $t=0.71$. As detailed in \cite{1906.08516}, we choose for our cosmological constant $\Lambda_{asymp}=1.025$. The parameters are chosen in such a way as to have an effective potential which is step-like with left plateau higher than the right one (i.e. we require $\frac{f_1^2}{f_2}>>\frac{M_1^2}{M_2}).$ While one may easily center the effective potential around $\phi=0$, this do not change qualitatively the observed results.

{\bf The case $b_0=0$}. 
 
We will work with the following parameters:

$M_0= -.03, M_1= 0.8, M_2= 0.01, \alpha= 2.4, p_u=10^{-85}, \chi_2= 1, f_1=5.7, f_2= 10^{-6}.$

For them, one can use two initial conditions: 

A. $a(0)=10^{-31}, \phi(0)=-1.8, \dot{\phi}(0)=0$

B. $a(0)=10^{-30}, \phi(0)=-3.8, \dot{\phi}(0)=0$

\begin{figure}[H]\centering{
\includegraphics[width=5cm]{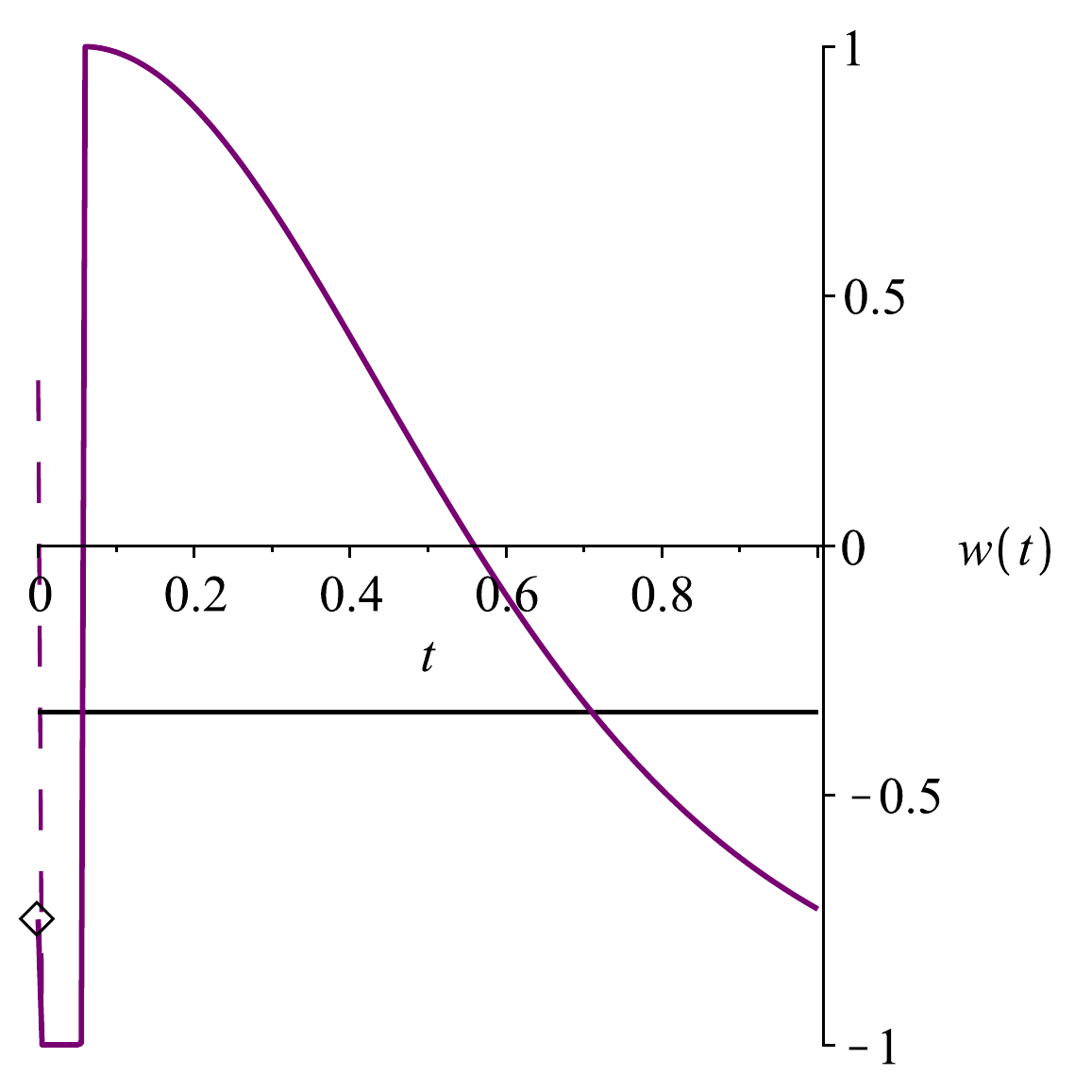} 
\includegraphics[width=5cm]{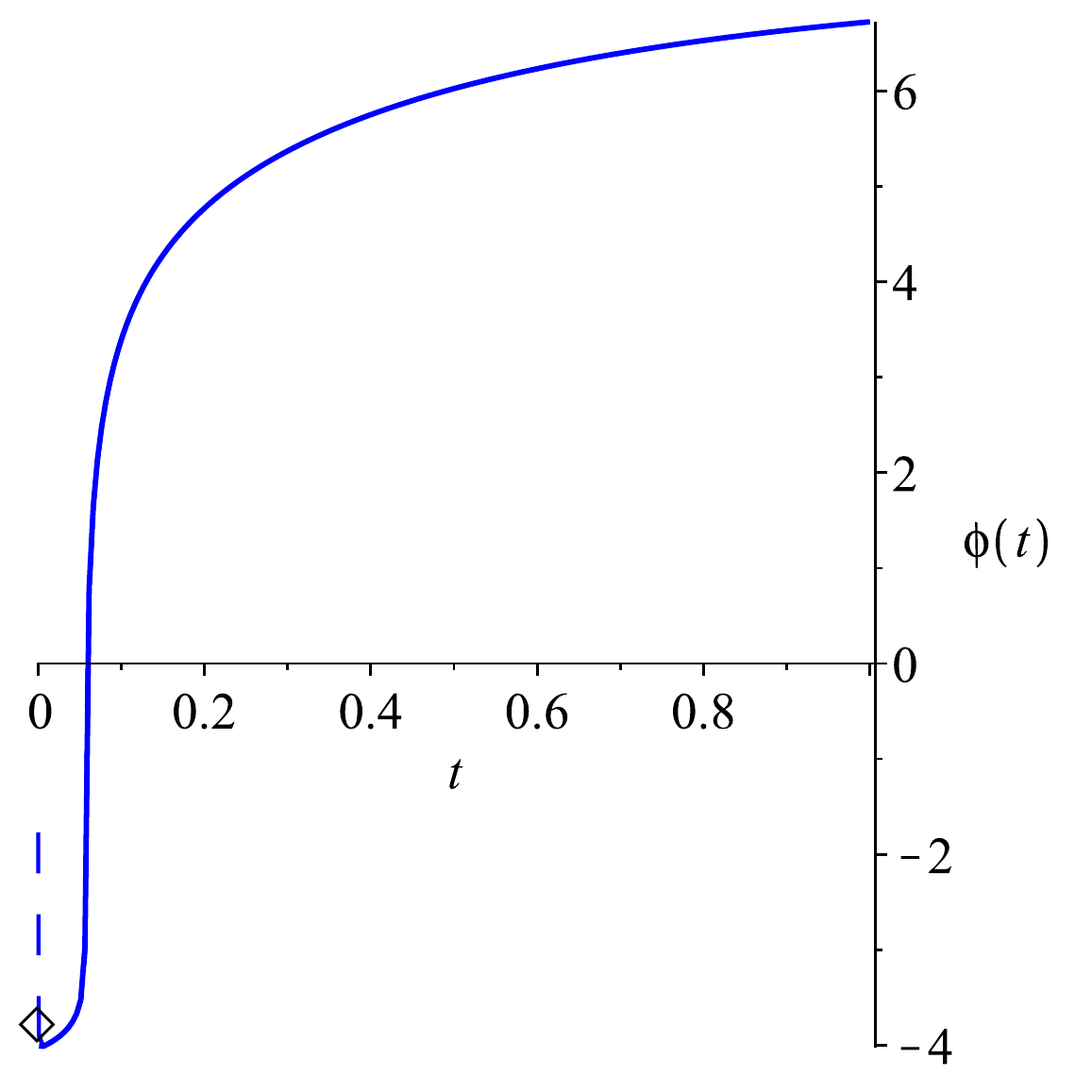}
\includegraphics[width=5cm]{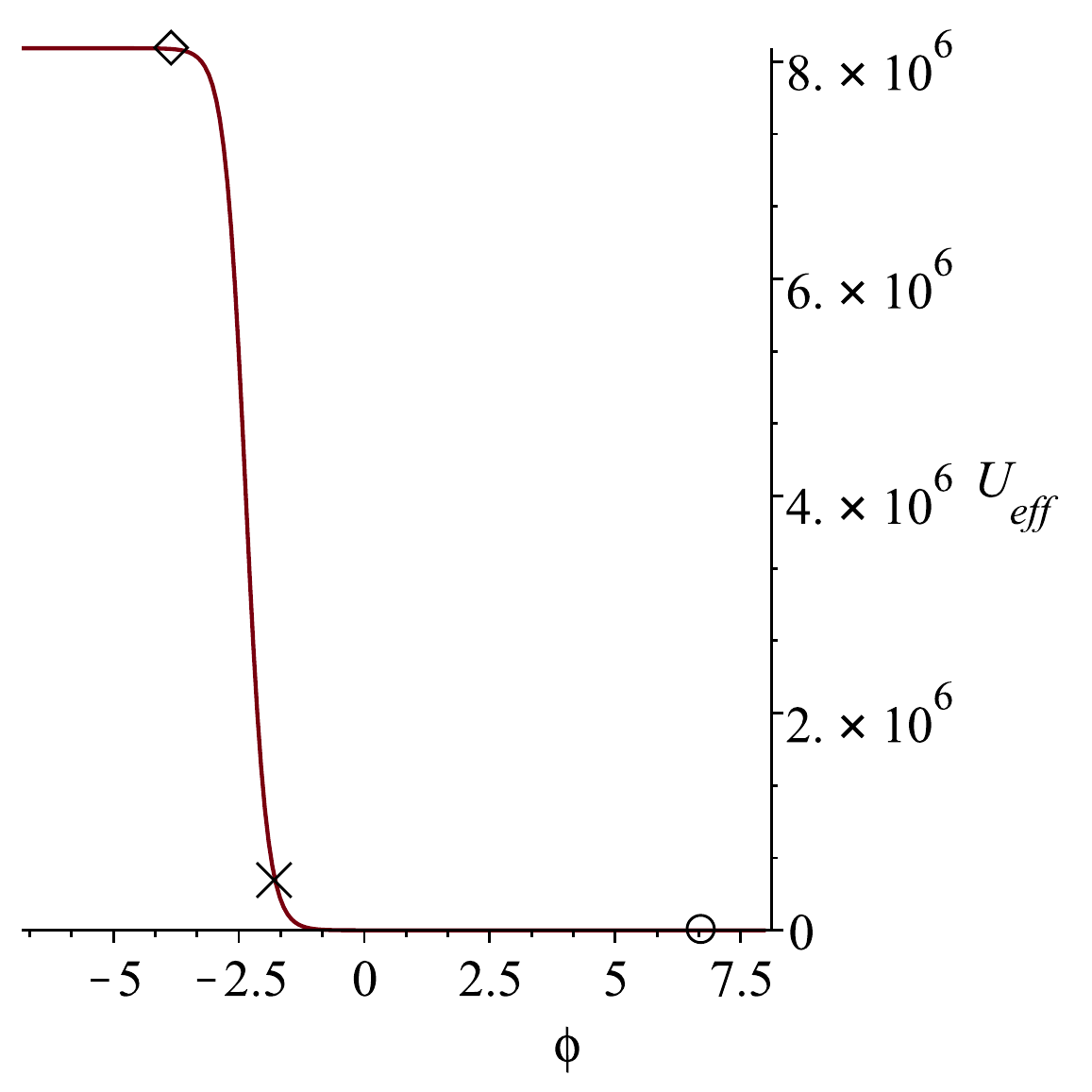}
}
\caption{From left to right:  
 the equation of state $w=p(t)/\rho(t)$, the inflaton field $\phi(t)$, and the effective potential $U_{eff}$. The dashed line corresponds to the case A, the solid -- to the case B. The cross and the diamond denote the start of the integration in the two cases, the circle - the final value for $\phi(t)$ }
\label{Fig1}
 \end{figure}

The plots of the relevant quantities for those two sub-cases are shown on Fig. \ref{Fig1}.  From the evolution of the EOS, one can see that in both case, we have a universe with 3 stages -- early inflation ($w\to-1$), matter-domination ($w<-1/3$) and late-time inflation ($w\to-1$). The two sub-cases match very closely in all time, except for the initial moments, when the first solution (A.) posses an ultra-relativistic stage ($w\to1/3$), marked on the plot with a dashed line. 

The evolution of the scalar field $\phi(t)$ in both cases is very similar, except for the first few time-steps of the integration. In both cases, it starts from certain value, it has a minimum and then it starts increasing. To understand better the movement of the inflaton, on the last plot we show the effective potential (same for both cases), with the starting points of our integration  marked with a cross for case A and with a diamond for case B. By tracing the movement of the scalar field, one can see that while case A starts much lower on the slope of the effective potential than case B, it reaches much higher on the effective potential. The inflaton field ``climbs up the slope'', i.e. it goes backwards instead of forward. This is much more pronounced in case A (climbing to to $\phi=-4$). In both cases, the inflaton stays on the slope of the potential  -- it does not reach the plateau characterized by $U_{eff}'(\phi_0) \to 0$ -- but it climbs to a much flatter part of the potential. This corresponds to what we have previously established in the general case, that the plateau is not accessible for the inflaton scalar field  \cite{1610.08368, 1801.07133, 1906.08516}. 

A phenomenon similar to ``climbing up the slope'' has already been observed in other inflationary theories. It has been proposed in \cite{1705.03023}  in a two-scalar fields model with a field space of a hyperbolic plane. Subsequently, the model has been generalised to more than 2 fields in \cite{1901.08603}. For this model, the second scalar field contributes to the so-called angular momentum. Instead of rolling down the potential, the scalar field would orbit the bottom of the potential until it has lost all its angular potential. This would lead to a prolonged inflation. According to the article, its perturbations are adiabatic and approximately scale invariant. While in our model, the inflaton does not orbit the bottom, but the top of the potential, it is still interesting to investigate the parallels between the two theories.

\begin{figure}[!ht]
\includegraphics[scale=0.25]{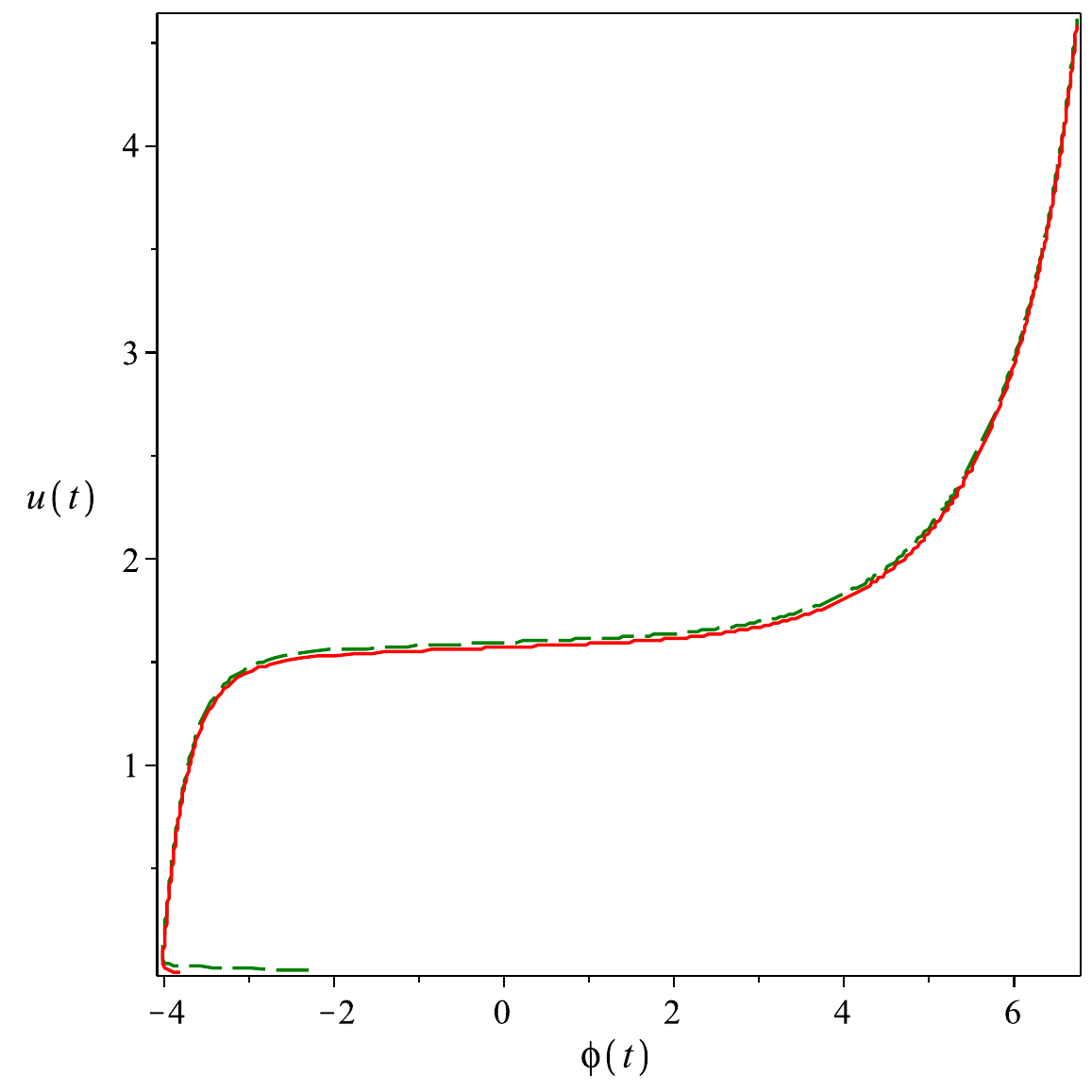}
\centering
\llap{\shortstack{%
        \includegraphics[scale=.1275]{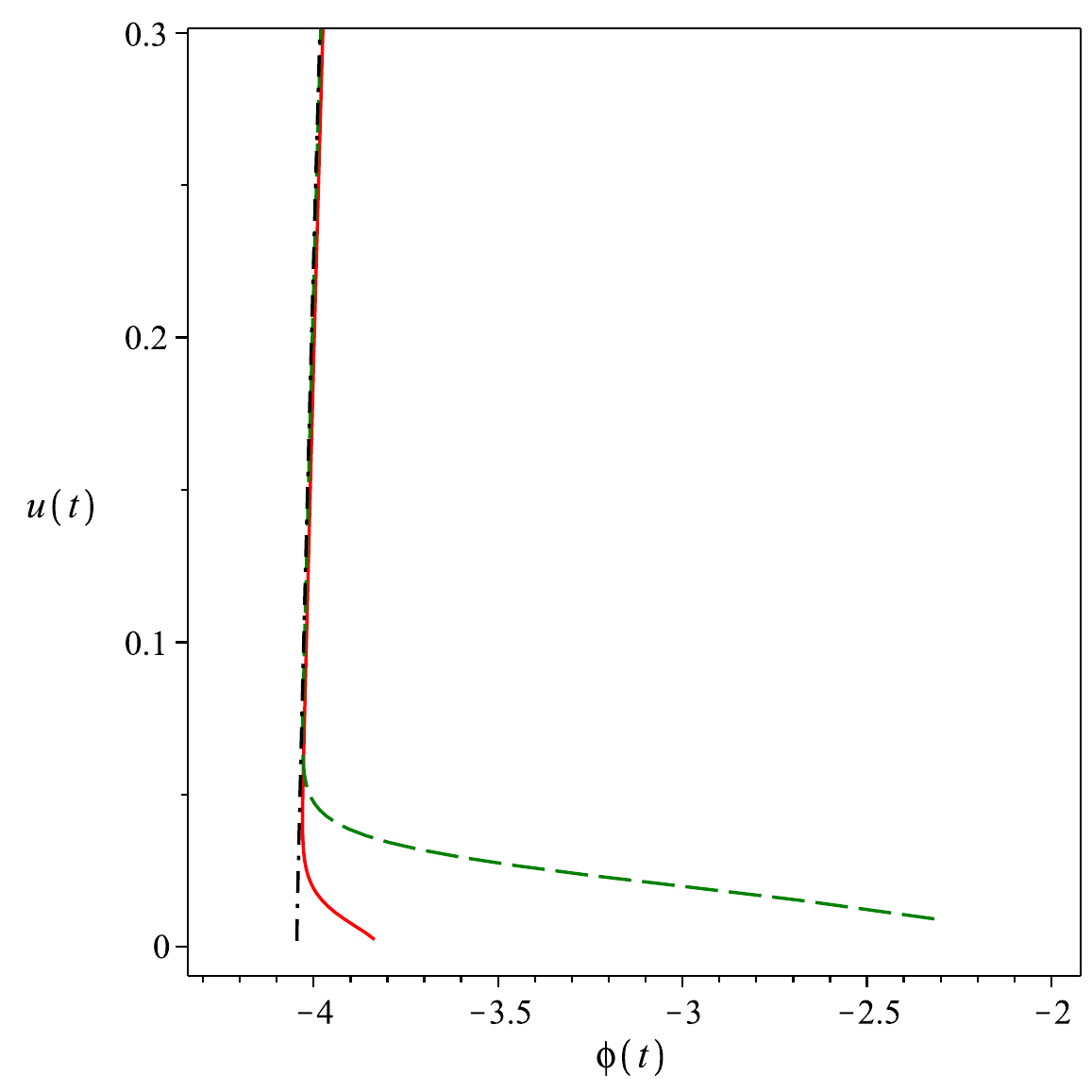}\\
        \rule{0ex}{0.85in}
      }
  \rule{0.65in}{0ex}} 
    \includegraphics[scale=0.25]{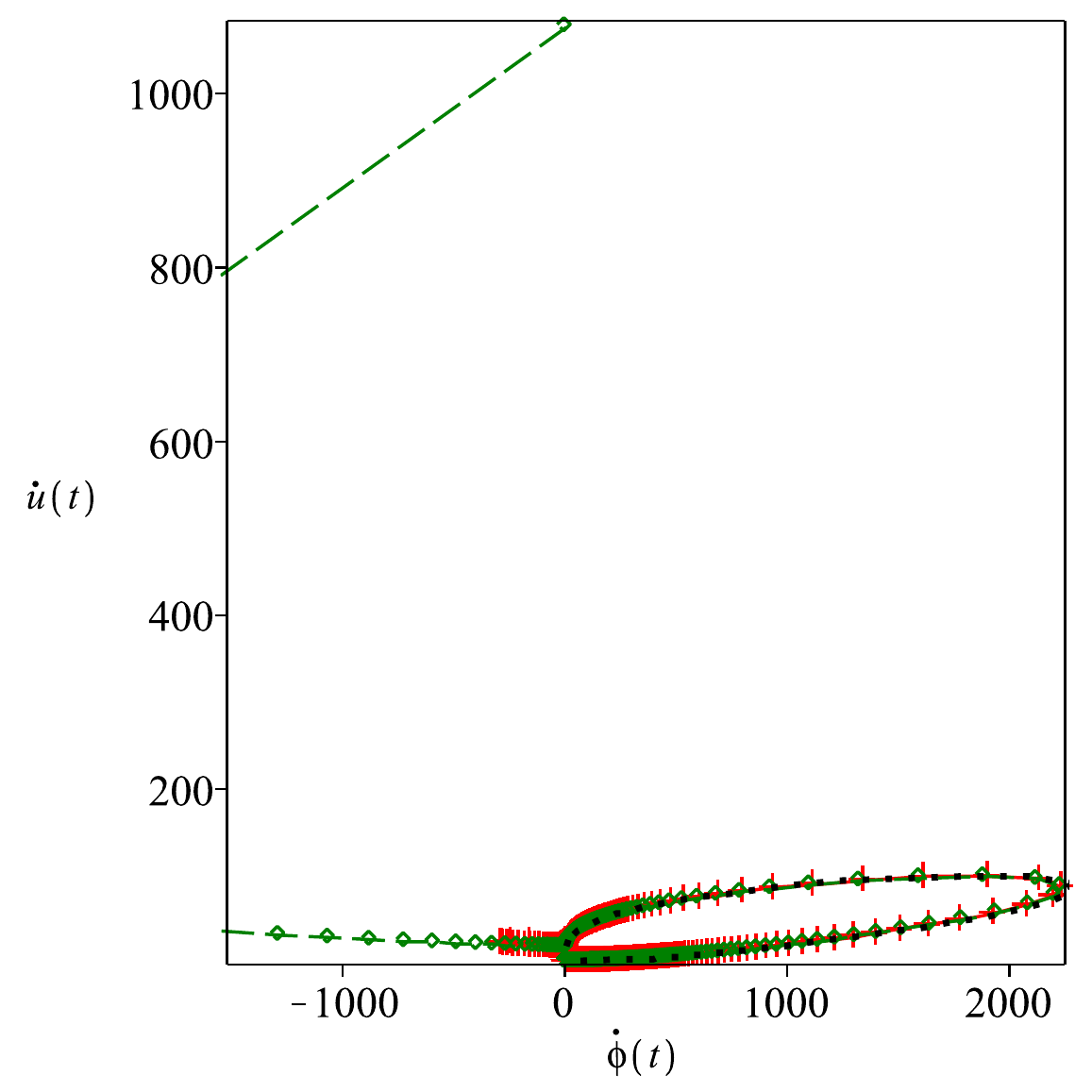}
\centering
\llap{\shortstack{%
        \includegraphics[scale=.1275]{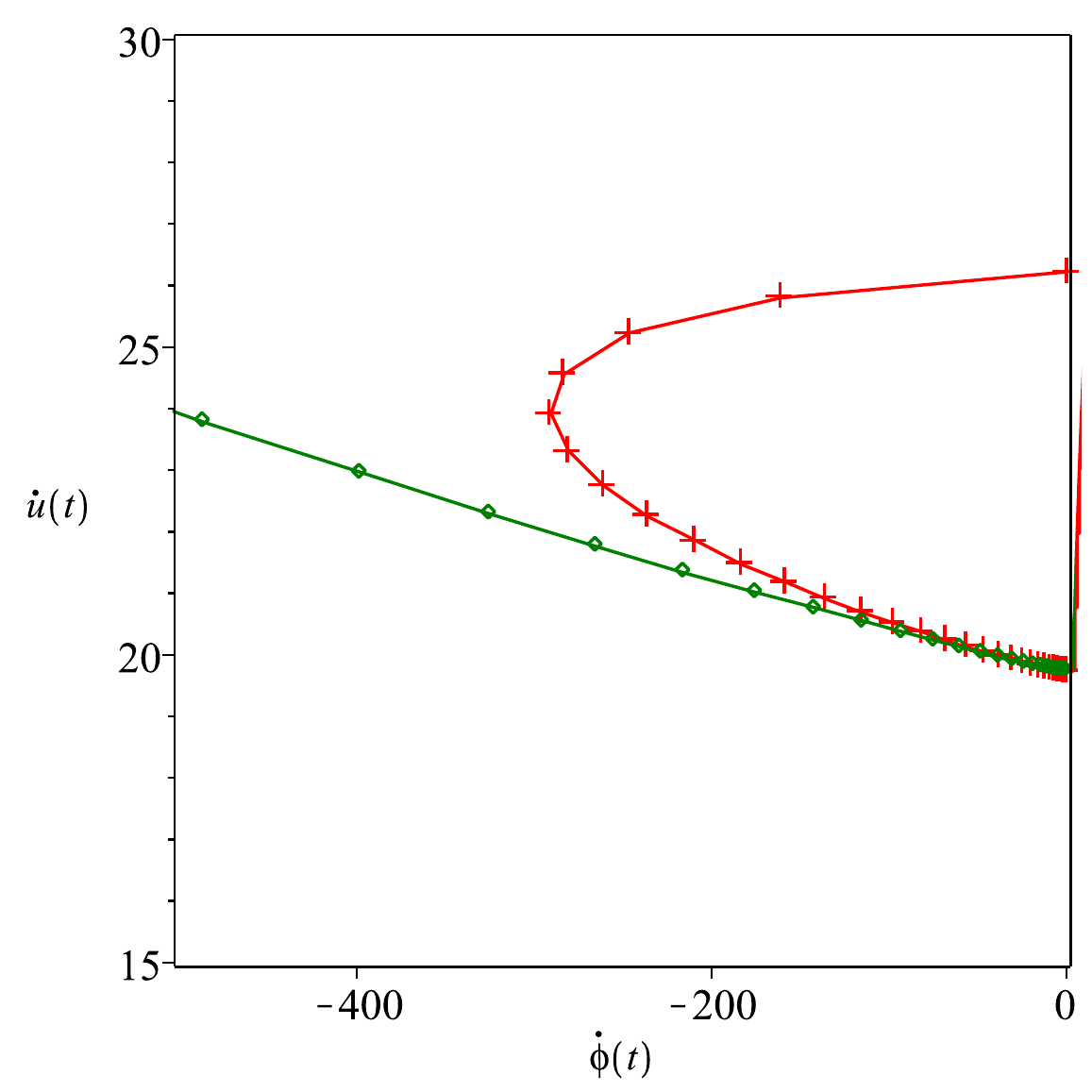}\\
        \rule{0ex}{0.77in}%
      }
  \rule{0.035in}{0ex}}
  \includegraphics[width=5cm]{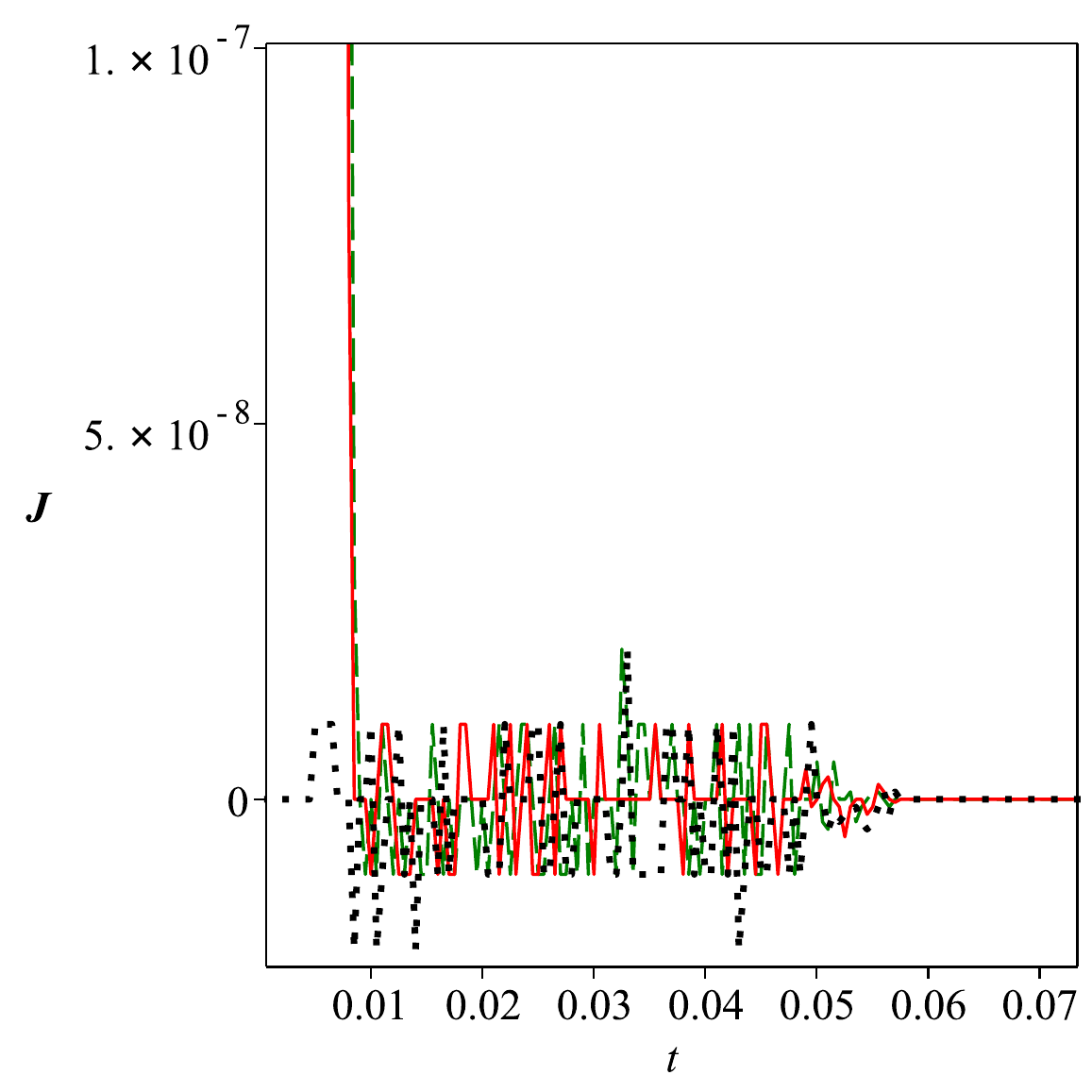}
\caption{The evolution of the two fields and their time derivatives. On the third plot we can see the so-called hyper-inflationary angular momentum. The color legend on the three plots is Case A: green dashed line, Case B: red solid line and $p_u=0$: the black dotted line }  
\label{Fig2}
\end{figure}

The Lagrangian of the hyperinflation model (assuming FRWL metric) in our notations is: 
 $$ L_{hyp}= \tilde{X} + \tilde{Y} f(\phi)  - V_{eff}(\phi). $$
 
The EOM for $u$ will lead to a conserved quantity $J_{hyp}(0)=a(t)^3J_{hyp}(t)$, where $J_{hyp}=f(\phi)v$ is the angular momentum. Inflation will happen until $J_{hyp}(t)\neq0$. In some regimes any angular perturbation may grow exponentially.

The comparison with the MMM can be done easily for $b_0=0$ when we put it in the form  $$ L_{MMM}= \tilde{X}+\tilde{Y}( -(V+M_1)+\tilde{Y^2}(\chi_2(U+M_2)-2M_0)).$$ Then the angular momentum will be: $J=-v\left( V+M_1\right)+v^3 (\chi_2(U(\phi)+M_2)-2M_0)$. Note that this corresponds exactly to the conserved current of the {\em dark fluid} in MMM (i.e. $p_u\to J_{hyp}(0)$):
\be
\partial_\mu\left(\sqrt{-\tilde{g}}\tilde{g}^{\mu\nu}\partial_\nu\tilde{u}\frac{\partial L^{(eff)}}{\partial \tilde{Y}}\right)=0.
\label{cons_curr}
\ee

To study the relationship between the two types of theories, we plot on Fig. \ref{Fig2} the scalar fields and their derivatives along with the angular momentum. The value for $u(t)$ is obtained after point-wise numerical integration using the modified Simpson’s rule applied to 10 000 points. One can see that the angular momentum starts very high for the cases when $p_u\neq0$, while it starts from approximately $0$ for $p_u=0$. In the three cases, the angular momentum starts oscillating around the zero while the inflation lasts and when inflation ends, it settle to zero.  Thus the angular momentum indeed traces the early inflation but it doesn't give information about the other stages trough which the evolution passes. The velocities of the two fields seem to exchange energy, except for the beginning of the evolution. Both of them demonstrate loops -- there is the late-time loop on the right ($\dot{\phi}>0$) common for both cases and the early-time one on the left ($\dot{\phi}<0$) which is much larger in the case A (the cut by the axes blue line), than it is in the case B (shown on the zoomed in plot with red). 

The inflation in the two cases happens for 
$\phi_A^I\in (-3.84,-2.21)$ and $\phi_B^I\in (-3.26,-2.29)$ which on the plots are the regions where $u$ is steeply rising while $\phi$ is almost constant. This correlates with the idea in hyperinflation theories that the scalar field orbits the potential. On the other hand, the matter domination happens while the darkon field is almost constant.  

We can conclude that the just like in the hyperinflation case, the observed epochs are born from the interplay between the two scalar fields and the exchange of energy between them. Inflation occurs while the inflaton remains approximately constant, i.e. $\dot{\phi}\approx0, \ddot{\phi}\approx0$, even though we start on the steep slope of the effective potential and not on its plateaus. Thus we have a slow-roll regime dynamically generated by the exchange of energy between the two scalar fields which sends the inflaton back on the more slowly varying upper part of the potential. The inflation ends when the angular momentum falls to almost zero and the system can no longer keep the inflaton on the flatter part of the potential and it starts rolling down the steep slope. 

We turn our attention to the number of e-folds, which can be calculated as $N=ln(a_f/a_i)$, where $a_i$ and $a_f$ are the beginning and the end of inflation, i.e we have removed the ultra-relativistic stage. The number of e-folds in case A is $N=68$, versus $N=63$ in case B. In both cases, it is enough to put the model in the error-bounds of observational expectations $N>65$. This is much higher than what we obtained in the general case ($b_0\neq0$), when the maximal value which we got was $N=22$. This effect seems to be mostly numerical as we will discuss later.

{\bf The case $b_0=0, p_u=0$}. 
The second special case which we consider is the one with $b_0=0$ and $p_u=0$. This case is particularly interesting because it removes partially the initial singularity from our equations and it allows us to study the Universe evolution when we do not have a Big Bang. In this case, the velocity of the darkon field is Eq. \ref{v_c0} and the inflaton equation is Eq. \ref{infl_eq}. Basically in this case, we have removed the evolution before $t=10^{-3}$ and we deal with much simplified equations of motion. The numerical evolution of the parameters follows very closely the ones shown on Fig. \ref{Fig1} for case B differing only that now the EOS  starts from $w(0)=-1$, while the inflaton field starts form its minimum shown on Fig. \ref{Fig1} ($\phi_0=-4.045$) and it increases monotonously afterwards. For this reasons, we omit showing it on Fig. \ref{Fig1}, and we add it only to the zoomed in figure of the dependence $u(\phi)$ on Fig. \ref{Fig2}, where it's shown with a black dash-dotted line. As for the second plot ($\dot{u}(t)(\dot{\phi}(t))$ the $p_u=0$ case has only the loop on the right, coinciding very closely with case B. For third plot, the difference is negligible, again closely resembling case B, only this time $J$ decreases monotonously, i.e. there is no inflexion point at the beginning. The most notable difference in this case is the lack of the ``climbing up the slope'' in the movement of the inflaton -- the scalar field just rolls down the slope as expected. This means that it is the dark charge what generates the "climbing up the slope" phenomenon.  Note that while, we have removed the singularity from $v(t)$, there is still the much weaker dependence on $a(0)$ coming from the inflaton equation itself. Because of this, different $a(0)$ will still affect the numerical solutions. 

{\bf Study of the parameters}

\begin{figure}[H]\centering{
\includegraphics[width=5cm]{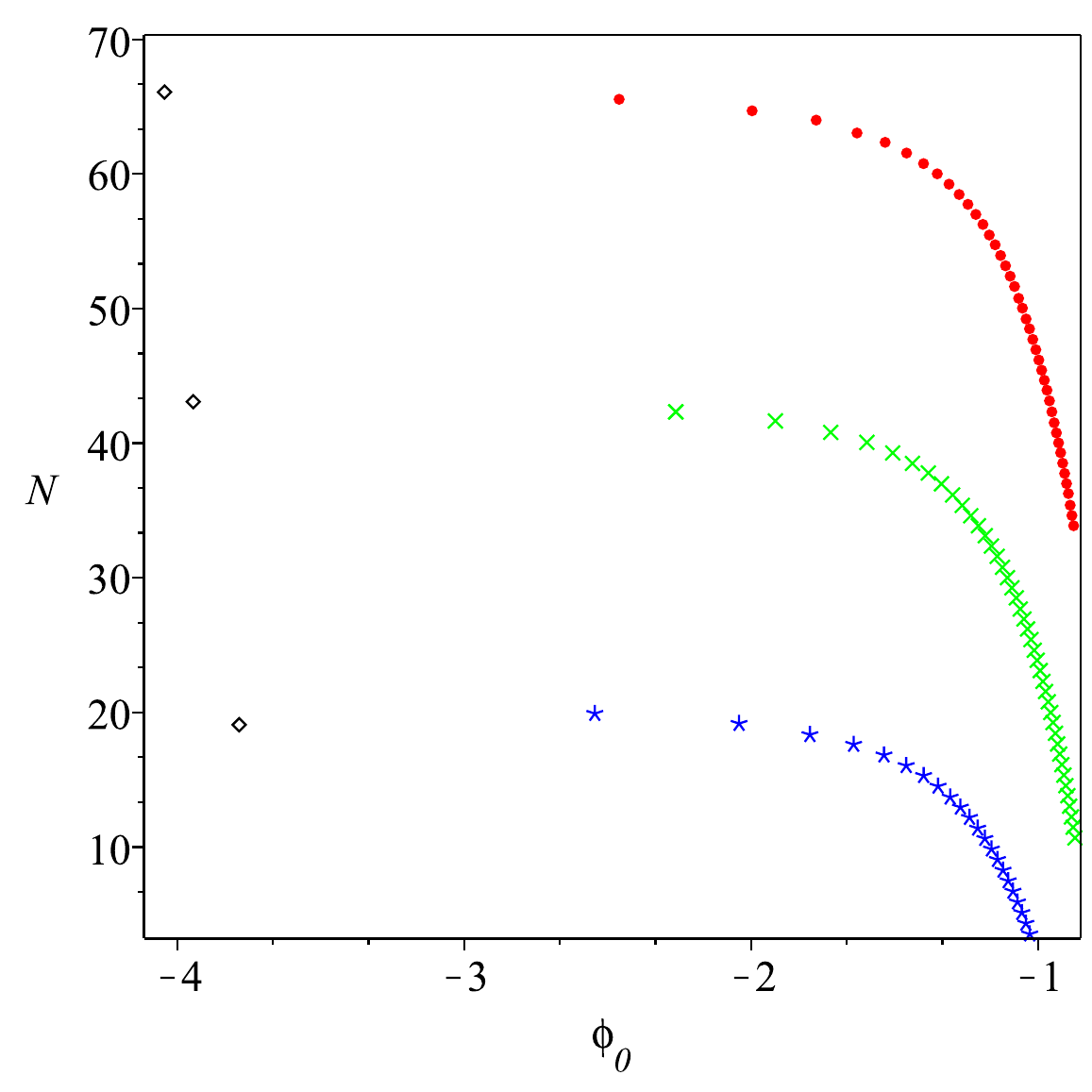} 
\includegraphics[width=5cm]{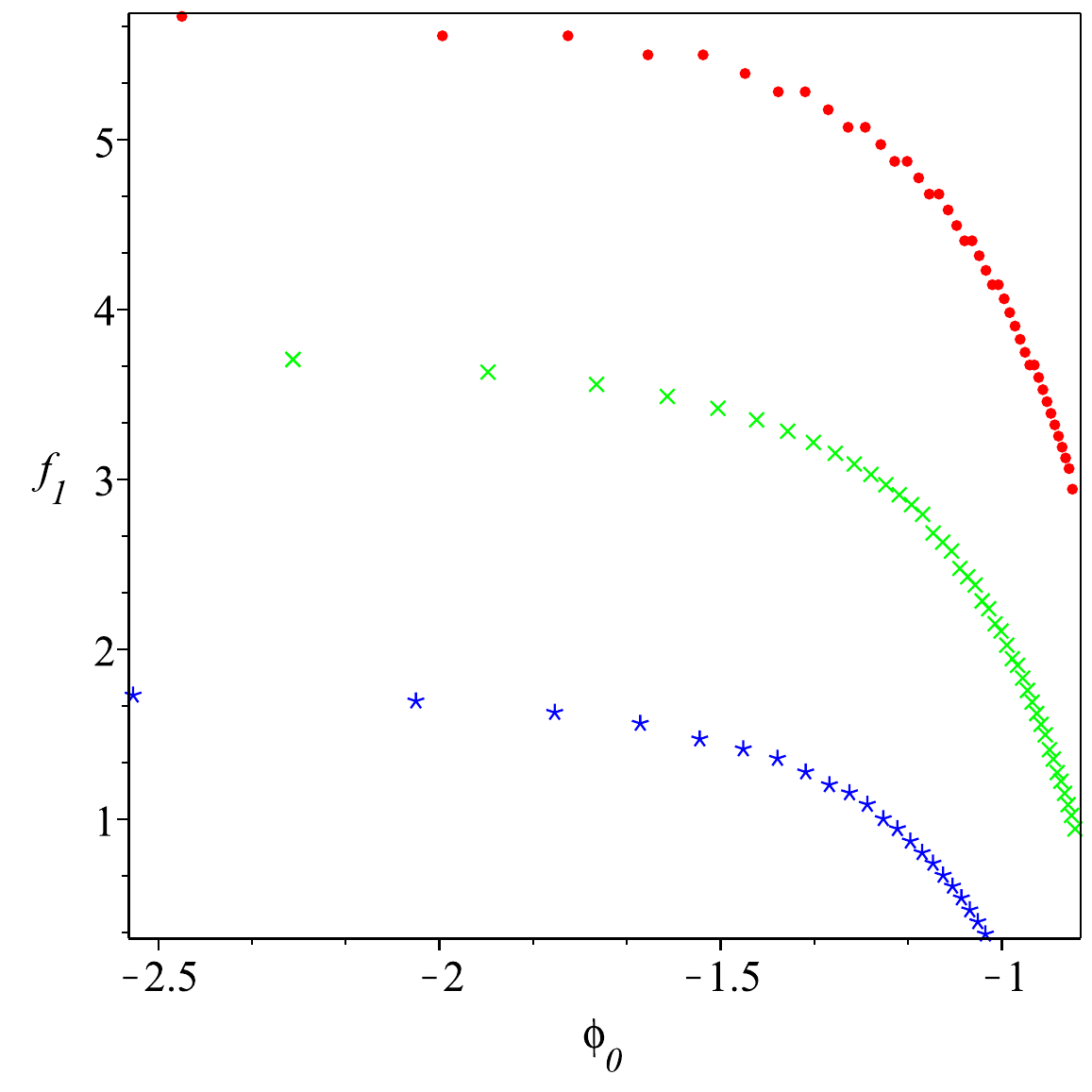}
\includegraphics[width=5cm]{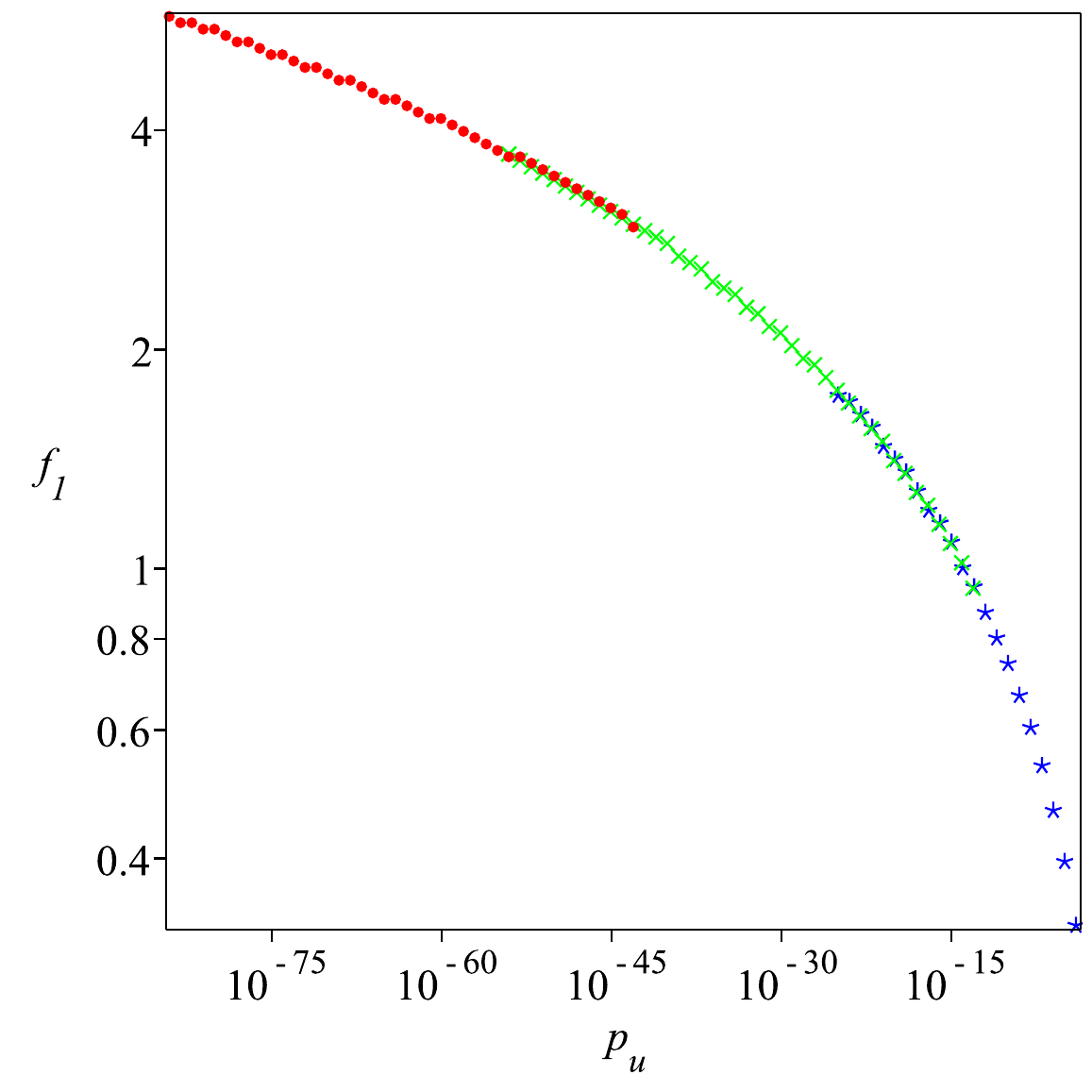}\\
a \hskip 5cm b \hskip 5cm c
}
\caption{On the panels we compare 3 values of the initial condition for $a(0)=10^{-10}, 10^{-20}, 10^{-30}$ denoted with asterisks, diagonal crosses and diamonds accordingly. We plot:
 a) the dependence of the number of e-folds from the starting point on the slope $\phi_0$, b) $f_1(\phi_0)$ c) $f_1(p_u)$ for the 3 different cases. }
 \label{Fig3}
 \end{figure}
In our previous works, we made the claim that the number of e-folds depends on the starting position of the integration and on the proximity of $a(0)$ to $0$. To study this phenomenon we wrote a code which automatically searches for solutions of the equations of motion, fulfilling our normalizations. This allowed us to study a much wider set of parameters by varying $a(0)$, $\phi(0), f_1, p_u$ to get the normalization $a(1)=1$ and $\ddot{a}(0.71)=0$. The results are shown on Fig. \ref{Fig3}, where we plot 3 sets of points corresponding to 3 different initial conditions: $a(0)=\{10^{-10}, 10^{-20}, 10^{-30}\}$.  The number of e-fold (Fig. \ref{Fig3} a))clearly grows with the decrease of the initial value of $a(0)$. This is apparently due to the initial singularity, but notably, not the one in $v(t)$, 
but the one in the inflaton equation itself. One can see this by noting the position of the black diamonds which correspond to the limit case $p_u=0$. This seems to confirm that indeed the problem of the too weak inflation numerically is due to the initial condition for $a(0)$ and that close enough to the singularity of the equations, one can get arbitrarily large number of e-folds. This seems as a numerical instability, but one must note that we do not know the initial conditions of our universe and thus we know only the minimal number of e-folds needed to produce our universe. Also since  all our solutions are normalized to $a(1)=1$, the Universe itself does not grow bigger, and the duration of the inflation does not grow for larger number of e-folds.

On Fig. \ref{Fig3} b) we have shown the relation $f_1(\phi_0)$ and on Fig. \ref{Fig3} c) the dependence $f_1(p_u)$. Surprisingly in the latter case, the values of different initial conditions $a(0), \phi_0$ fall on the same curve, which show that this relation is independent from the initial conditions. We recall that both $f_1$ and $p_u$ are physical quantities, one of them comes from the potential term of one of the inflaton Lagrangians while $p_u$ is the conserved dark charge. 

{\bf The speed of sound}

\begin{figure}[H]\centering{
\includegraphics[width=5cm]{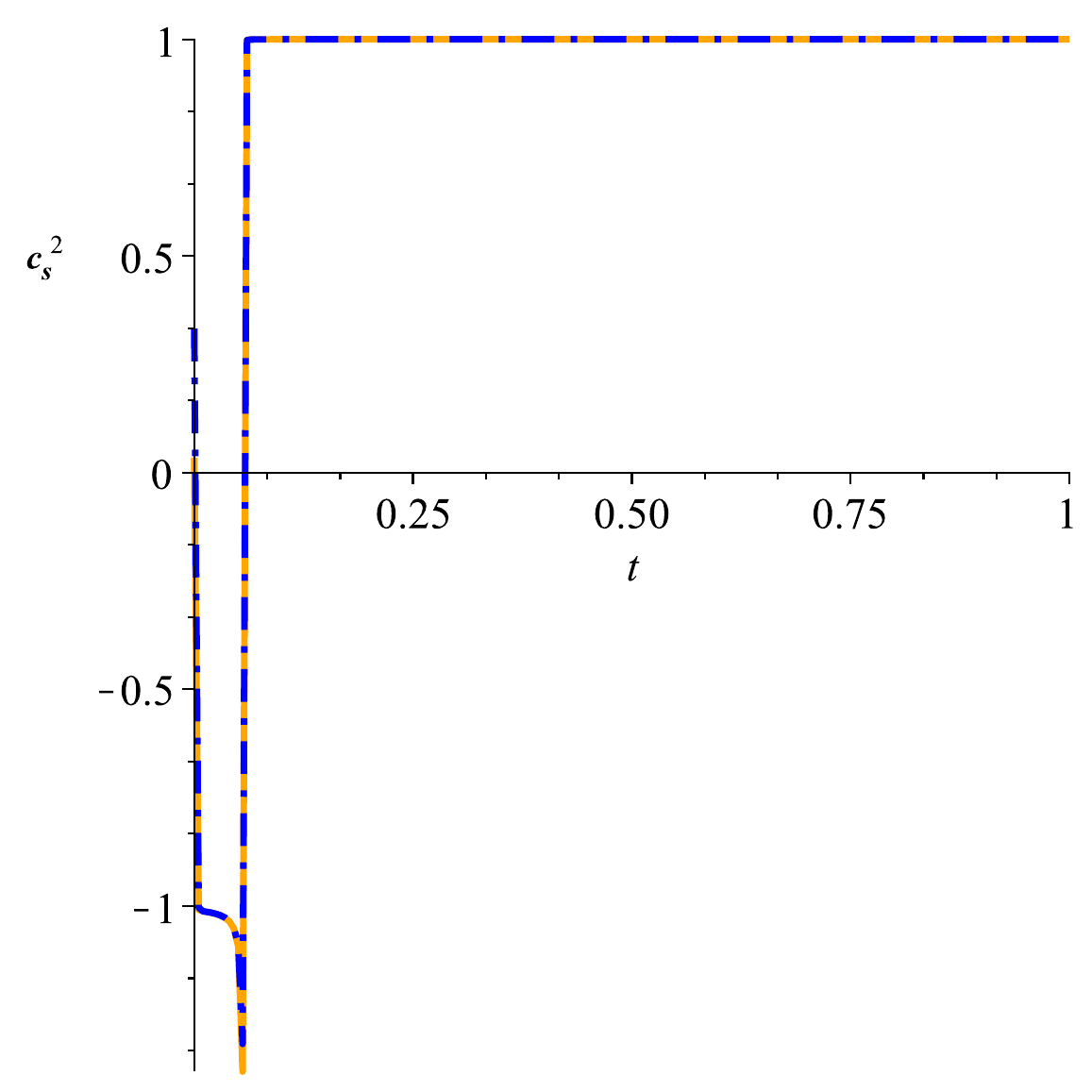} 
\includegraphics[width=5cm]{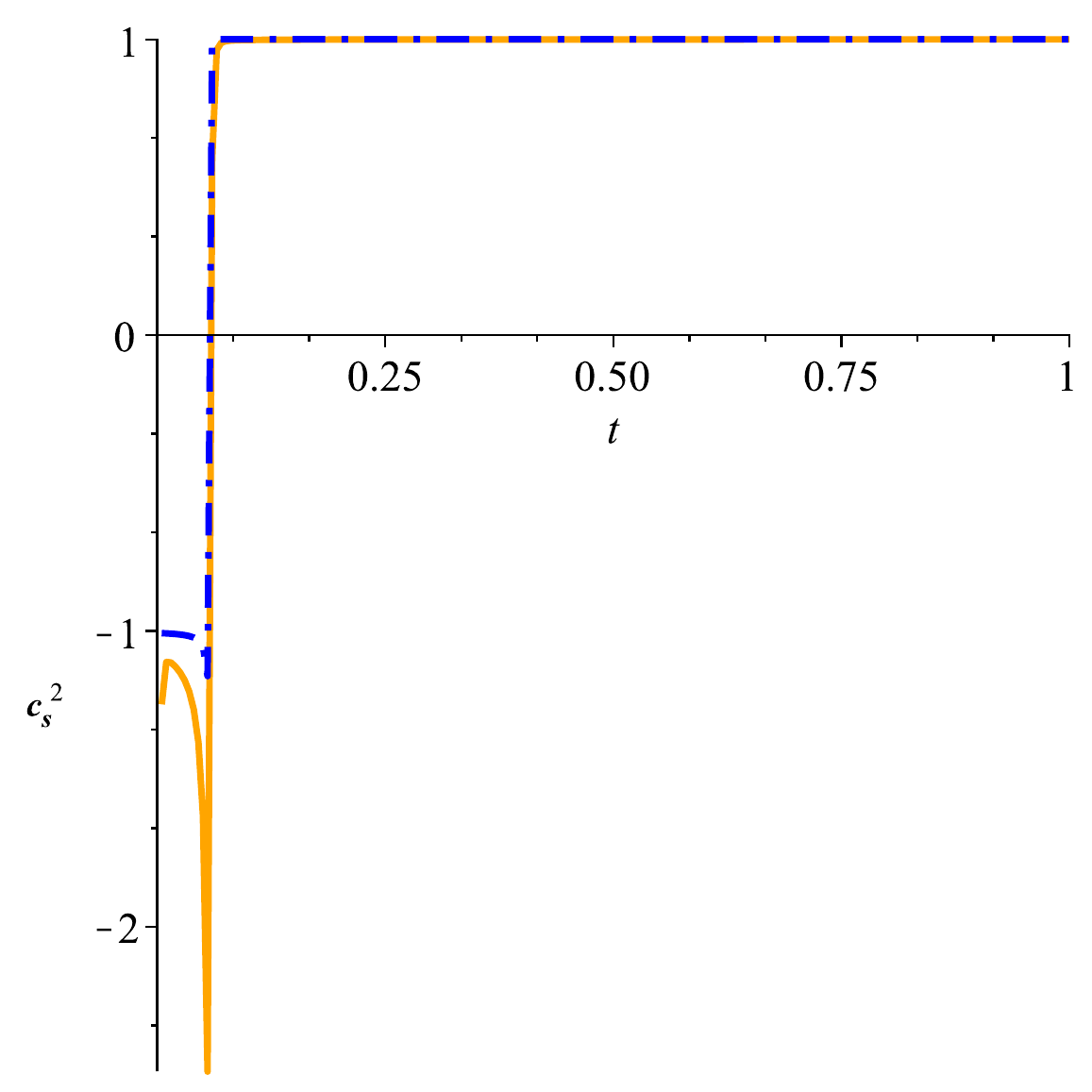}\\
a \hskip 5cm 
}
\caption{On the panels are the speed of sound of the two main case  
 a)  $b_0=0, p_u\neq0, a(0)=10^{-30}$ (case A) denoted with solid yellow line , case (B) denoted with blue dash-dot line
 b) $b_0=0, p_u=0, a(0)=10^{-5}$ denoted with blue, dashed line, $b_0=0, p_u=0, a(0)=10^{-45}$ -- with yellow, solid line}
 \label{Fig4} 
 \end{figure}

Finally, we would like to discuss the speed of sound in the two special cases. 
We have plotted the so called adiabatic sound speed, $c{_s}^2{_a}=\frac{\dot{p}}{\dot{\rho}} $ 
on Fig. \ref{Fig4} for a)  the $p_u\neq0$ cases from above, and on b) the $p_u=0$ cases for two different starting points $a(0)=10^{-5}$ and $a(0)=10^{-45}$.

Despite the different initial conditions and parameters in the 4 cases, it doesn't seem possible to avoid the negative  region. Even in the case (B) where there is a positive initial speed of sound it still reverts to $-1$ during inflation. This is well-known property of perfect fluid dark energy models and it is considered related to the difference between the adiabatic speed of sound defined above (generated by pressure perturbations) and the actual speed of propagation of the perturbation, the phase speed (generated by entropic perturbations).

 {Calculating the phase speed of sound in theories with non-canonical kinetic terms of two scalar fields is not trivial. A generalization for a wide range of multifield theories of the form $L=P(X^{IJ},\phi^K)$ can be found in \cite{0806.0336, 0801.1085}. According to \cite{0801.1085}, the perturbations along the field-space trajectory move with the single-field $c_{s, phase}$, while the orthogonal ones move with the speed of light. The effective speed of sound is defined as $c_{s, phase}^2=p,_X/\rho,_X=\frac{P_X}{P_X+2XP_{XX}}$, where ${}_{,X}$ is the derivative with respect to $X$.For Lagrangians with cannonical kinetic terms, i.e. that are a sum of scalar fields of the form $X=G_{IJ}X^IX^J$, where $X=\partial_i (\phi^\alpha)$ and $G_{IJ}$ are functions of the fields, the effective sound speed is $1$ \cite{Gao:2008dt}. In our case, the contribution of the non-linear term can be evaluated using the formula for the phase speed to 1/3, which means that overall effective sound speed is going to be $c_s^2\lesssim 1$ but positive}. The phase speed in the special case $b_0=0, p_u=0$, on-shell, coincides with that of a standard one scalar field theory, i.e. it equals 1 \cite{0809.3518}.

There are a number of theories in which the adiabatic speed of sound shows non-standard behavior but the effective speed of sound (the phase speed) gives a scale on which perturbations may be dampened, the so called effective sound horizon (see \cite{9801234, 0404494}). For example, in quintessence, the speed of sound is imaginary, while in $k-essence$ theories,  $c_s^2>1$ and thus perturbations can travel faster than light. In \cite{1707.03885}, quintessence with non-minimal derivative coupling to gravity has been shown to suffer from both superluminal perturbations and Laplacian (gradient) instability -- $c_s^2<0$.  In \cite{1711.05196} the authors have studied variable dark energy speed of sound and have found that $c_s^{DE}\neq 1$ when the model has non-canonical kinetic term. In a study \cite{1804.11279, 1805.12563} inspired by the sidetracked inflation, but generalized to models allowing an effective single field theory, the imaginary sound speed leads to exponentially increasing and decreasing modes (instead of positive and negative modes). The exponentially growing fluctuation becomes constant after the sound Hubble crossing, so they are named transient tachionic instability. For constant roll inflation with multi scalar fields \cite{guerrero2020constant} it has been found that entropy perturbations become null, while \cite{Hohmann_2019} find that in teleparallel gravity theories all waves propagate with the speed of light. An interesting study \cite{0307100} shows that to fit WMAP data,  $c_s^2<0.04$. Similarly, in \cite{1004.5509}, the authors predict that dark energy clustering is more efficient when $c_s\to0$. In our case, we can see that while the adiabatic speed of light indeed is variable and imaginary during inflation, the phase speed does not imply instabilities. We leave the complete investigation of the perturbation in the special cases and in the general case for future works.

\section{Conclusions} 
In this article, we have studied two special cases of the multi-measure model of Guendelman--Nissimov--Pacheva, for which one can decouple the kinetic terms of the two scalar fields in the effective Lagrangian. Those special cases preserve the evolution of the universe having the known 3 stages without any further constraints. We show numerically that one is able to get the necessary number of e-folds and thus it is possible to obtain a strong-enough early inflation. This along with the fact that the model does not modify the speed of light makes it a viable candidate for description of the universe. 

Our most interesting result is the connection of the MMM with the hyperinflationary models in which the ‘centrifugal force’ of a field orbiting the hyperbolic plane of the two fields leads to a prolonged inflation. We have studied how this movement of the scalar fields with respect to each other is related to the equation of state of the universe and to the so-called angular momentum of the model. We have seen that during the early inflationary epoch we have a dynamically induced slow-roll period in which the effective potential varies more slowly thus allowing the application of slow-roll approximations. That means that all the results of the slow-roll approximation are valid even if we do not start on the plateau of the potential as usually assumed. 

Finally, we show that while numerically the adiabatic speed of sound becomes imaginary during inflation, the phase speed calculated for the model is close to or equal to the speed of light.

\vspace{6pt}

{\it{\textbf{Acknowledgments}}} 
The work is supported by the Bulgarian National Science Fund for support via research grants DN 08-17, DN-$18/1$, KP-06-N$38/11$. We have received partial support from European COST actions CA15117 and CA18108. It is a pleasure to thank Emil Nissimov, Svetlana Pacheva, Michail Stoilov and David Benisty for the discussions.


%

\bibliographystyle{apsrev4-2}
\bibliography{DStaicova_SpecialCases}
\end{document}